\newif\ifsubmissionversion
\submissionversionfalse

\ifsubmissionversion
	\documentclass{ar-1col}
\else
	\documentclass[11pt,]{article}
\fi

\usepackage{graphicx}
\usepackage{subcaption}
\usepackage{amsmath,amssymb}
\usepackage{amsfonts}
\usepackage{ulem}
\usepackage[comma]{natbib}
\usepackage{float}
\usepackage[hidelinks]{hyperref}

\ifsubmissionversion
\jname{Xxxx. Xxx. Xxx. Xxx.}
\jvol{AA}
\jyear{YYYY}
\doi{10.1146/((please add article doi))}
\fi

	\graphicspath{{figs/}}

\renewcommand{\emph}[1]{{\textit{#1}}}
\newcommand{\E}{\mathbb{E}}

\newcommand{\details}[1]{}

\begin{document}


\title{Spatial Population Genetics: It's About Time}

\ifsubmissionversion
\else

\author{Gideon S. Bradburd$^{1a}$ and Peter L. Ralph$^2$}
\maketitle

\noindent\textsuperscript{1}Ecology, Evolutionary Biology, and Behavior Group, Department of Integrative Biology, Michigan State University, East Lansing, MI, USA, 48824\\

\noindent\textsuperscript{2}Institute of Ecology and Evolution, Departments of Mathematics and Biology, University of Oregon, Eugene, OR 97403\\

\noindent\textsuperscript{a}bradburd@msu.edu\\
\fi

\begin{abstract}
    Many questions that we have about the history and dynamics of organisms
    have a geographical component:
    How many are there, and where do they live?
    How do they move and interbreed across the landscape?
    How were they moving a thousand years ago,
    and where were the ancestors of a particular individual alive today?
    Answers to these questions can have profound consequences
    for our understanding of history, ecology, and the evolutionary process.
    In this review, we discuss how geographic aspects of the 
    distribution, movement, and reproduction of organisms
    are reflected in their pedigree across space and time.
    Because the structure of the pedigree is what determines 
    patterns of relatedness in modern genetic variation,
    our aim is to thus provide intuition for how these processes
    leave an imprint in genetic data.
    We also highlight some current methods and gaps 
    in the statistical toolbox of spatial population genetics.
\end{abstract}

\ifsubmissionversion
\else
\newpage
\fi

\tableofcontents

\ifsubmissionversion
\else
	\newpage
\fi

\section{Introduction}

The field of population genetics is shaped by a continuing conversation
between theory, methods, and data.
We design experiments and collect data
with the methods we will use to analyze them in mind, 
and those methods are based on theory
developed to explain observations from data.
With some exceptions,
especially the work of Sewall Wright \citeyearpar{Wright1940,Wright1943,wright1946isolation}
and Gustave Mal\'ecot \citeyearpar{malecot},
much of the early body of theory was focused on
developing expectations for discrete populations.
This focus was informed by early empirical datasets,
most of which
\citep[again, with exceptions, like][]{Dobzhansky_Wright1943, dobzhansky1947}
were well-described as discrete populations --
like samples of flies in a vial \citep{lewontin1974}.
In turn, 
the statistics used to analyze these datasets --
e.g., empirical measurements of $F_{ST}$ \citep{Wright1951}
or tests of Hardy-Weinberg equilibrium \citep{hardy1908,weinberg1908} --
were defined based on expectations in discrete, well-mixed populations.
In part, this was because much more detailed mathematical predictions 
were possible from such simplified models.
The focus of theory and methods on discrete populations in turn influenced
future sampling designs.

To what degree does reality match this picture
of clearly delimited yet randomly mating populations?
Often, the answer is: not well.
Organisms live, move, reproduce, and die,
connected by a vast pedigree that is anchored in time and space.
But, it is difficult to collect data and develop theory and methods
that capture this complexity.
Sampling and genotyping efforts are limited by access, time, and money.
Developing theory to describe evolutionary dynamics
without the simplifying assumption of random mating
is more difficult.
For example, without discrete populations,
many commonly-estimated quantities in population genetics -- 
migration rate ($m_{ij}$), 
effective population size ($N_e$),
admixture proportion
--
are poorly defined.
A long history of theoretical work describing geographical populations
\citep[e.g.,][]{fisher1937wave,haldane1948theory,slatkin1973geneflow,nagylaki1975conditions,sawyer1976branching,barton1979dynamics}
has not had that strong an effect on empirical work \citep[but see][]{landscape_genomics_review}.
However, modern advances in acquiring both genomic and geospatial data
is making it more possible -- and, necessary -- to explicitly model geography.
Today's large genomic datasets
are facilitating the estimation of the timing and extent of shared ancestry 
at a much finer geographic and temporal resolution than was previously possible
\citep{Li_Durbin2011,Palamara_2012,Harris_Nielsen_2013,ralph2013geography}.
And, advances in theory in continuous space 
\citep{felsenstein1975pain,barton1995genealogies,barton-depaulis-etheridge, barton2010newmodel, hallatschek2011noisy, barton2010modelling, Barton2013},
datasets of unprecedented geographical scale
\citep[e.g.,][]{POBI, Aguillon2017deconstructing, Shaffer195743},
new computational tools for simulating spatial models \citep{haller2018forward,haller2019treesequence},
and new statistical paradigms for modeling those data 
\citep{petkova2016visualizing, ringbauer2017inferring, ringbauer2018estimating, conStruct, alasadi2018estimating}
are together bringing an understanding of the geographic distribution of genetic variation into reach.

The goal of this review
is to frame some of the fundamental questions in spatial population genetics,
and to highlight ways that new datasets and statistical methods 
can provide novel insights into these questions.
In the process, 
we hope to provide an introduction for empirical researchers
to the field.
We focus on a small number of foundational questions 
about the biology and history of the organisms we study:
where they are; how they move; where their ancestors were;
how those quantities have changed over time, 
and whether there are groups of them.
We discuss each question
as a spatial population genetic problem, 
explored using the \emph{spatial pedigree}.
It is our hope that by framing our discussion around the spatial pedigree, 
we can build intuition for --
and spur new developments in --
the study of spatial population genetics.

\section{The spatial pedigree}

\begin{figure}
    \centering
         \includegraphics[width=\linewidth]{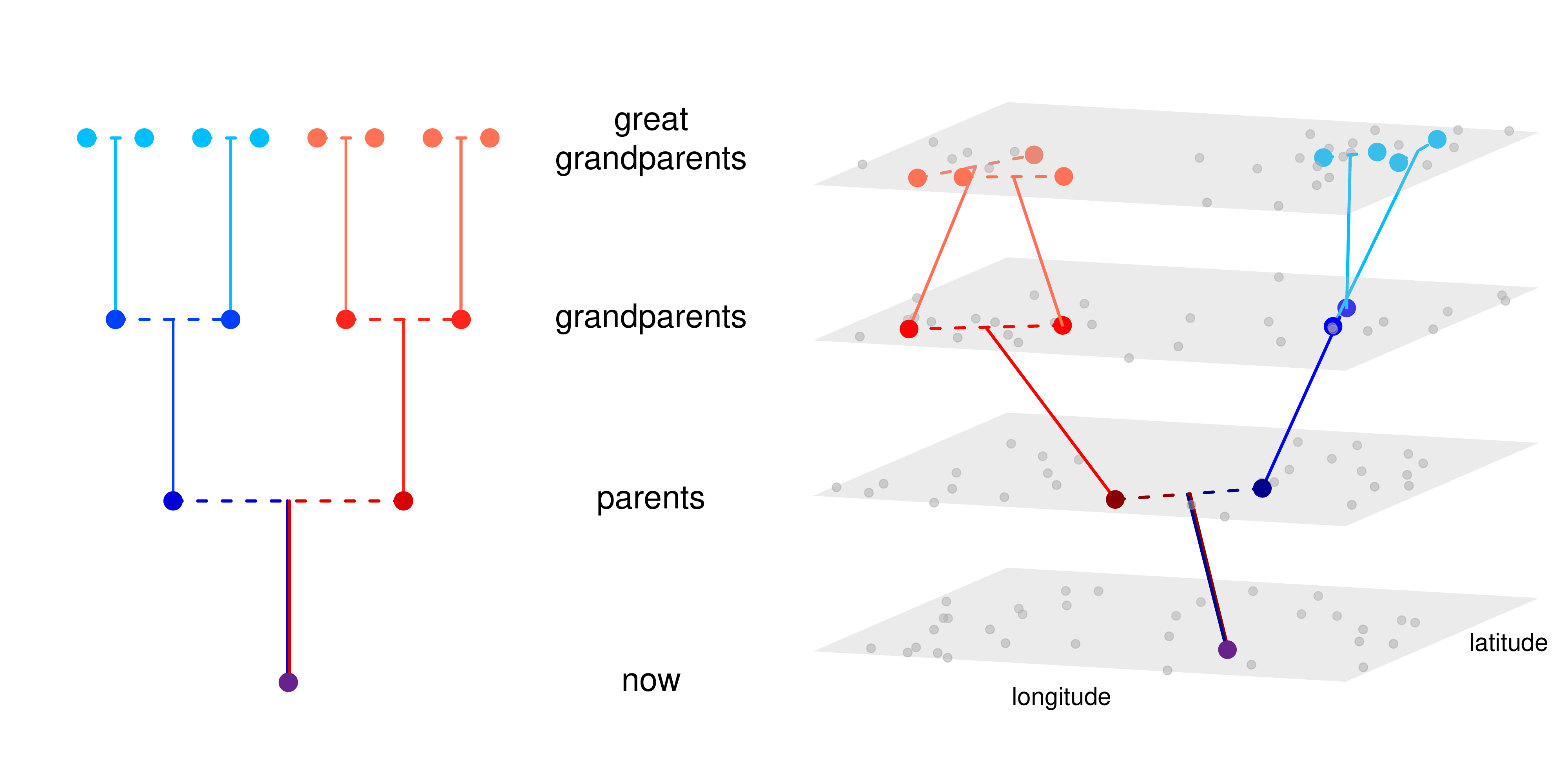}
        \caption{
		An example pedigree (left) of a focal individual sampled in the modern day, 
	   	placed in its geographic context to make the spatial pedigree (right).
		Dashed lines denote matings, and solid lines denote parentage, 
		with red hues for the maternal ancestors, 
		and blue hues for the paternal ancestors.
		In the spatial pedigree, 
		each plane represents a sampled region in a discrete (non-overlapping) generation,
		and each dot shows the birth location of an individual.
		The pedigree of the focal individual is highlighted 
		back through time and across space.
        }
        \label{spatial_pedigree}
\end{figure}

The spatial pedigree 
encodes parent-child relationships of all individuals in a population,
both today and back through time,
along with their geographic positions (Figure \ref{spatial_pedigree}).
If we had complete knowledge of this vast pedigree,
we would know many evolutionary quantities
that we otherwise would have to estimate from data,
such as the true relatedness between any pair of individuals,
or each individual's realized fitness.
If we wanted to characterize the dispersal behavior of a species,
we could simply count up the distances between offspring and their parents.
Or, if we wanted to to know whether a particular allele
was selected for in a given environment,
we could compare the mean fitness of all individuals in that environment with that allele
to that of those without the allele.
The ``location'' of an individual can be a slippery concept,
depending on the biology of the species in question
(as can the concept of ``individual'').
For instance, 
where in space does a barnacle hitchhiking on a whale live, 
or a huge, clonal stand of aspen?
To reduce ambiguity,
we will think of the location of an individual as the place it was
born, hatched, or germinated.

In practice, we can never be certain of the exact spatial pedigree.
Even if we could census every living individual in a species
and establish the pedigree relationships between them all,
we still would not know all relationships between -- 
and locations of -- 
individuals in previous, unsampled generations.
And, in a highly dispersive species,
or a species with a highly dispersive gamete,
there may be only a weak relationship between observed location and birth location.
Despite these complexities,
the idea of the spatial pedigree is useful
as a conceptual framework for spatial population genetics.
Although it may be difficult or impossible
to directly infer the spatial pedigree,
we can still catch glimpses of it through the window of today's genomes.

\subsection{Peeking at the spatial pedigree}
Early empirical population genetic datasets were confined to 
investigations of a small number of individuals 
genotyped at a small number of loci in the genome.
Many of the statistics measured from early data, 
such as $\pi$, $\theta$, or $F_{ST}$,
rely on averages of allele frequencies
at a modest number of loci,
and thus reflect mostly deep structure at a few gene genealogies.
These statistics are informative about 
the historical average of processes like gene flow acting over long time scales, 
but are relatively uninformative about processes in the recent past, 
or, more generally, about heterogeneity over time.
However, the ability to genotype a significant proportion of a population 
and generate whole genome sequence data for many individuals 
is opening the door to inference of the recent pedigree 
via the identification of close relatives in a sample.
In turn, we can use this pedigree, 
along with geographic information from the sampled individuals, 
to learn about processes shaping the pedigree in the recent past.

\subsection{Simulating spatial pedigrees}

It has only recently become possible to simulate whole genomes
of reasonably large populations
evolving across continuous geographic landscapes.
Below, 
we illustrate concepts with 
simulations 
produced using SLiM v3.1 \citep{haller2018forward},
which enables rapid simulation of complex geographic demographies.
We recorded the genealogical history along entire chromosomes
of entire simulated populations
as a \emph{tree sequence} \citep{haller2019treesequence}.
In most simulations, we retained \emph{all} individuals alive at any time during the simulation
in the tree sequence, allowing us to summarize not only genetic relatedness
but also genealogical relatedness 
(e.g., to identify ancestors from whom an individual has not inherited any parts of their genome).
We used tools from \texttt{pyslim} and \texttt{tskit} \citep{kelleher2018efficient} 
to extract relatedness information,
and plotted the results in \texttt{python} with \texttt{matplotlib} \citep{hunter2007matplotlib}.
Using SLiM's spatial interaction capabilities,
we maintained stable local population densities by increasing mortality rates of individuals
in regions of high density,
and chose parameter values to generate relatively strong spatial structure
(dispersal distance between 1 and 4 units of distance 
and mean population density 5 individuals per unit area).
Our code is available at \url{https://github.com/gbradburd/spgr}.
This new ability to not only simulate large spatial populations,
but also to record entire chromosome-scale genealogies as well as spatial pedigrees,
has great promise for the field moving forward.

\subsection{Estimating ``effective'' parameters from the spatial pedigree}

\begin{quote}
    \textit{
    On one hand, every single one of my ancestors going back billions of years
    has managed to figure it out.
    On the other hand, that's the mother of all sampling biases.}
    \hfill \textit{-- Randall Munroe, xkcd:674}
\end{quote}

Inferences about the past
based on those artifacts that have survived to the present
are fraught with difficulties and caveats in any field, 
and population genetics is no exception.
The ancestors of modern-day genomes are \emph{not} an unbiased sample
from the past population -- 
they are biased towards individuals with more descendants today.
It is obvious that we cannot learn anything from modern-day genomes
about populations that died out entirely, 
but there are more subtle biases.
For example, if we trace back the path along which a random individual today 
has inherited a random bit of genome,
the chance that a particular individual living at some point in the past is an ancestor
is proportional to the amount of genetic material that individual has contributed to the modern population.
If we are looking one generation back, 
this is proportional how many offspring they had, i.e., their fitness.
If we are looking a long time in the past
(in practice, more than 10 or 20 generations \citep{BartonEtheridge2011fitness}),
this is their \emph{long-term fitness}.
Individuals from some point in the past who have many descendants today
may or may not look like random samples from the population at the time.
Different methods query different time periods of the pedigree,
and are thus subject to these effects to different degrees.

A familiar example of the limits of inference 
is the difference between ``census'' and ``effective'' population size \citep{Wright_1931,Charlesworth2009}.
More subtle is the difference between dispersal distance ($\sigma$) and
``effective dispersal distance'' ($\sigma_e$) \citep[reviewed in][]{Cayuela2018demographic}.
Both are mean distances between parent and offspring:
$\sigma$ is the average observed from, say, the modern population,
but $\sigma_e$ is
the mean distance between parent and offspring along a lineage back through the spatial pedigree.
The difference arises because
lineages along which genetic material is passed do not give an unbiased sample of all dispersal events;
they are weighted by their contribution to the population.
For instance, if there is strong local intra-specific competition,
individuals that dispersed farther from their parents (and their siblings)
may have been more likely to reproduce and be ancestors of today's samples,
which would make $\sigma_e > \sigma$.
On the other hand, if there is local adaptation,
individuals may be dispersing to habitats dissimilar to that of their parents
and are unlikely to transmit genetic material to the next generation
because they are maladapted to the local environment,
and so $\sigma_e < \sigma$
\citep[for a review, see][]{wangbradburd2014}.

\section{Things we want to know}

\subsection{Where are they?}

Perhaps the first things we might use
to describe a geographically distributed population
are an estimate of the total population size
and a map of population density across space.
Quantifying abundance and how it varies
can be crucial, e.g., for understanding
what habitat to prioritize for conservation \citep{zipkin2018synthesizing}, 
which regions have higher population carrying capacities than others \citep{roughgarden1974}, 
or whether a particular habitat is a demographic source or sink 
\citep{pulliam1988sources}.
The best measure of population density
would be a count of the number of individuals occurring in various geographic regions, 
and the best measure of total population size would be the sum of the counts across all regions.
How close can population genetics get us to estimating these quantities by indirect means?
Some fundamental limitations are clear:
for instance, if we only sample adults,
we cannot hope to estimate the proportion of juveniles that die before adulthood.
None of our sampled genetic material inherits from these unlucky individuals,
so they are invisible as we move back through the pedigree.
In addition, it is tricky to separate the inference of population density 
from that of dispersal, 
as the rate of dispersal to or from a region of interest 
governs whether it makes sense to consider the pedigree in that region on its own.
Partly because of this issue, 
it is also difficult to estimate total size from a sample of a 
non-panmictic population, 
and so it is more intuitive to discuss local density first, 
then scale up to the problem of total population size.
The spatial pedigree contains a record of both total and regional population sizes, 
and we can use the genomes transmitted through it to learn about these quantities.

\begin{figure}	
    \centering
    \begin{subfigure}{0.95\textwidth}
        \centering
        \includegraphics[width=\linewidth]{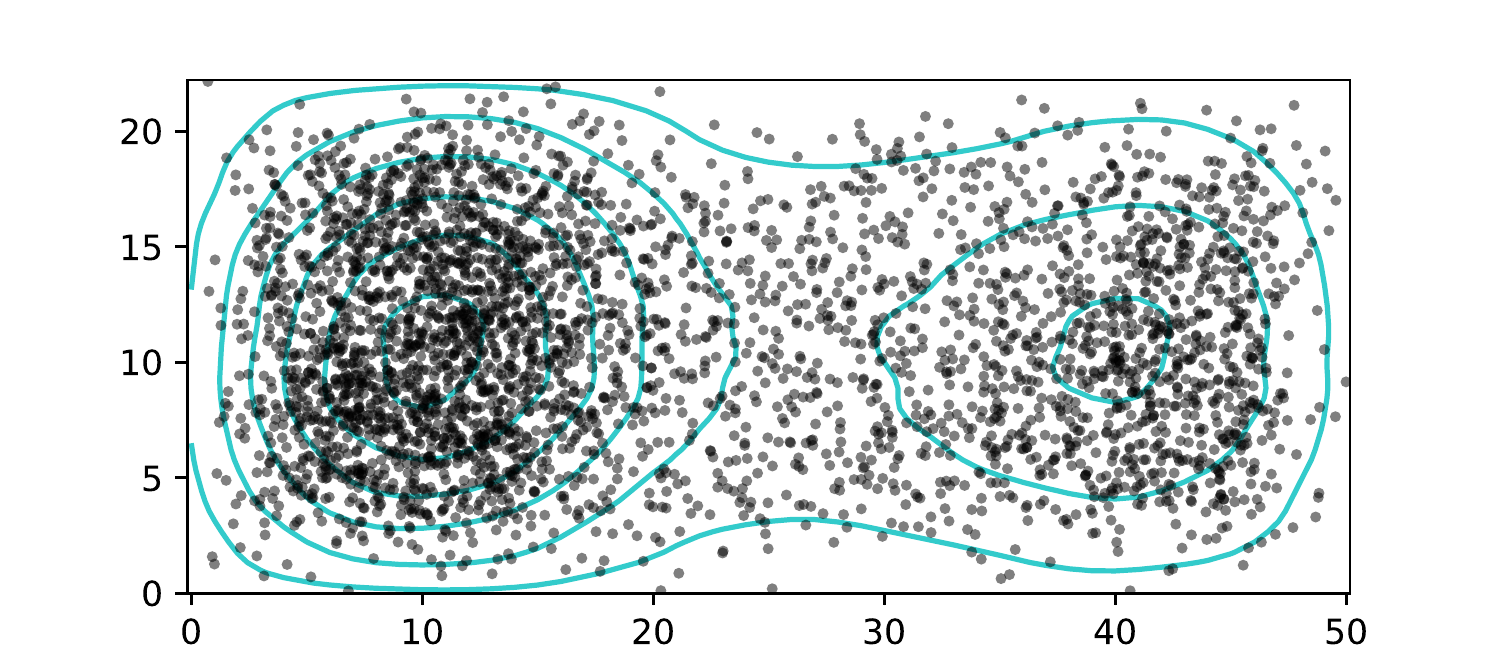}
        \caption{map of simulation scenario}
        \label{valley_map}
    \end{subfigure}
    \begin{subfigure}{0.95\textwidth}
        \centering
        \includegraphics[width=\linewidth]{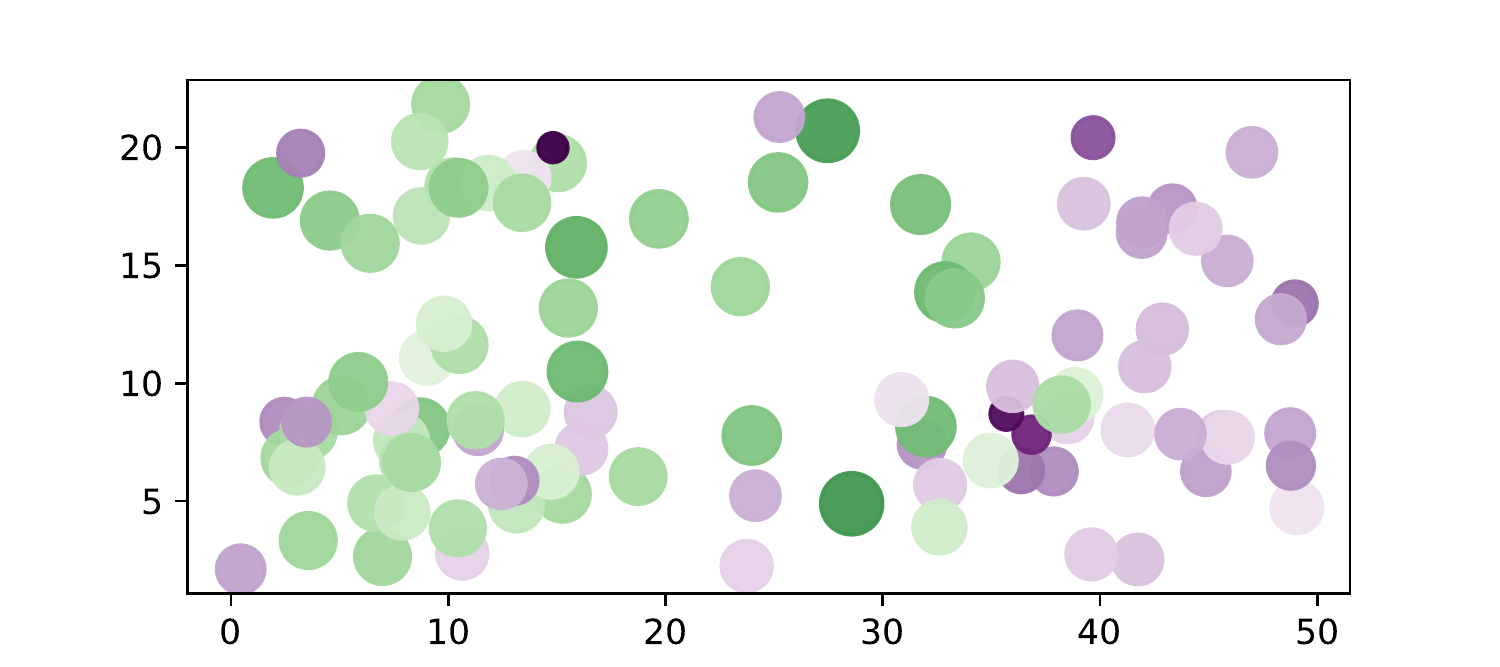}
        \caption{individual heterozygosity}
        \label{valley_het}
    \end{subfigure}
        \caption{
            \textbf{(a)} Map of population density of the simulated population scenario
            we use for examples throughout.
            Populations have local density-dependent regulation
            within two valleys with high local carrying capacity (indicated by contours),
            separated and surrounded by areas with lower carrying capacity.
            The western valley has a carrying capacity about twice as large as that of the eastern valley.
            \textbf{(b)} Heterozygosity of a sample of 100 individuals across the range.
            Circle sizes are proportional to heterozygosity,
            and circle color indicates relative values
            (green: above the mean; white: mean; purple: below the mean).
            Low heterozygosities occur on the edges of the range,
            and, because there is a strong barrier between the two valleys,
            mean heterozygosity is decreased by about 40\% in the lower-density eastern valley.
		}
        \label{pop_density}
\end{figure}

\subsubsection{Population density}

Genetic information about population sizes or densities most clearly
comes from numbers and recency of shared ancestors, 
i.e., the rate at which \emph{coalescences} between distinct lineages
occur as one moves back through the pedigree.
The relationship between population density and coalescence is, in principle, straightforward:
with fewer possible ancestors,
individuals' common ancestors tend to be more recent.

Some calculations may be helpful to further build intuition
(although the details of the calculations are not essential to what follows).
Consider the problem of inferring the total size of the previous generation
in a randomly-mating population
from the number of siblings found in a random sample of size $n$.
There are $x (x-1) / 2$ sibling pairs in a family with $x$ offspring,
so the probability that, 
in a population of size $N$, 
a given pair of individuals are siblings in that family is $x(x-1)/N(N-1)$.
Averaging across families, the probability that a given pair of individuals
are siblings (in any family)
is therefore $\E[X (X-1)] / (N-1)$, where $X$ is the number of offspring in a randomly chosen family.
This implies that in a sample of size $n$,
$n(n-1)/2$ divided by the number of observed sibling pairs 
provides an estimate of \emph{effective population size}, 
$N_e = N/\E[X(X-1)]$.
A similar argument appears in \citet{mohle2003cpd}; 
also note this is \emph{inbreeding} effective population size
(see \citet{wang2016prediction} for a review of different types of ``effective population size'').
In other words, we should be able to use relatedness to estimate population size
up to the factor $\E[X(X-1)]$, which depends on demographic details. 
This same logic can be applied locally:
the (inbreeding) effective population density
near a location $x$, denoted $\rho_e(x)$,
can be defined as the inverse of the rate of appearance of sibling pairs
near $x$, per unit of geographic area and per generation \citep{barton-depaulis-etheridge}.
\details{
    Let $P$ denote the probability that two diploids sampled from the population are siblings, and
    let $Q$be the probability that randomly sampled chromosomes from two randomly sampled diploids inherit from the same parental chromosome at a randomly chosen locus.
    Claim: Assuming monogamy (no half-sibs), $1/P$ is ``diploid inbreeding $N_e$'', 
    defined (in \citet{ewens2004mpg}) to be $1/(2Q)$.  
    (Note the caveats about diploidy do not appear in Ewens.)
    Proof: Under monogamy, the chance a random locus on random chromosomes from the two diploids are IBD from a common parent is $1/2$, so $Q = P/2$.
}

We can extend this line of reasoning to see that 
the local density of close relatives
should reflect local population size.
The chance that two geographic neighbors actually \emph{are} siblings 
decreases with the number of ``possible nearby parents'':
\citet{wright1946isolation} quantified this with the
\emph{neighborhood size}, 
$N_\text{loc} = 4 \pi \rho_e \sigma_e^2$,
where $\pi$ is the mathematical constant,
$\rho_e$ is the effective population density,
and $\sigma_e$ is the effective dispersal distance.
Each pair of siblings gives a small amount of information
about population density within distance $\sigma_e$ of the pair.
The more distant the coalescent event relating a pair of individuals, 
the vaguer the information they provide about population density,
because there is greater uncertainty 
both in the inference of their precise degree of relatedness from genetic data,
and in the location of their shared ancestors.
Because genetic estimates of population size 
derive from rates of coalescence,
they generally can only be used to estimate population sizes 
in the locations and times where those coalescences occur.

Relatives up to recent cousins can be identified in a large sample to varying degrees of certainty 
from genomic data \citep[reviewed in][]{wang2016prediction},
using multilocus genotypes \citep{nomura2008estimation,WaplesWaples2011,Wang_2014},
shared rare alleles \citep{NovembreSlatkin2009}, 
or long shared haplotypes \citep[e.g.,][]{li2014relationship}.

An alternative approach to estimating local density
is to measure the rate of genetic drift in an area.
To do this, one compares allele frequencies
in at least two samples collected in a region in different generations, 
either at different points in time,
or, in a species with overlapping generations, 
between contemporaneous individuals of different life stages, 
e.g., saplings and mature trees
\citep{WilliamsonSlatkin1999}.
The rate of drift is governed by, 
and therefore provides an estimate of, 
local population size; 
this is analogous to the \emph{variance effective size} of a population
\citep{ewens2004mpg,Charlesworth2009,wang2016prediction}.
Because these approaches are pinned to the recent past, 
either by the generations sampled or by the degree of closeness of relatives,
they should offer a more accurate estimate of recent population density than heterozygosity does, 
especially when density has changed over time.

\subsubsection{Total population size}

Estimates of the largest geographic scale -- total population size --
turn out to be sensitive to changes over the longest time scales.
Total effective population size is often estimated using measures of genetic diversity such as $\pi$ \citep{NeiLi1979,Tajima89},
the mean density of nucleotide differences between two randomly sampled chromosomes.
Because each such difference is the result of a mutation
sometime since the chromosomes' most recent common ancestor at that location on the genome,
$\pi$ is equal to twice the mean time to most recent common ancestor
within the sample, multiplied by the mutation rate.
In other words, if the per-nucleotide mutation rate is $\mu$,
and $2T$ is the mean length of a path through the pedigree
along which two samples have inherited a random bit of genome
from their most recent common ancestor,
then $\pi \approx 2 T \mu$.
The quantity $T$ itself is an interesting summary of the pedigree,
and depends on geography, sampling, and dispersal in complex ways.
What does it have to do with population size?
In a randomly mating, diploid population of constant effective size $N_e$,
the discussion above implies that $T = 2N_e$.
In practice, populations are rarely constant, and so this is viewed as an average
over the last few $N_e$ generations.

\subsubsection{Heterozygosity and population density}

We have seen that
local population size is reflected in the pedigree through rates of shared ancestry:
all else being equal, smaller populations should have more shared ancestry and 
therefore less overall genetic diversity.
To gain intuition, we can look at how genetic diversity on the smallest scale
-- heterozygosity --
behaves in a simulated example.
Consider two adjacent valleys, 
one densely populated, the other sparsely populated, 
but both equal in geographic area 
and separated by a barrier to dispersal (Figure \ref{pop_density}\subref{valley_map}).
In general, we expect pairs of individuals in the low density valley 
to have shared ancestors in the more recent past than 
pairs of individuals in the densely populated valley.
This happens simply because 
each group of siblings makes up a larger proportion 
of the more sparsely populated valley.
Indeed, individuals in the sparsely populated valley 
have more shared ancestors in the more recent past, 
so the two chromosomes sampled in each individual 
from that valley are more closely related,
i.e., have lower heterozygosity, as shown in Figure \ref{pop_density}\subref{valley_het}.

Heterozygosity correlates with population density in this example with a strong barrier,
but does it predict relative density more generally?
Differences in heterozygosity can be shaped by other forces.
For example, we might also expect a similar difference in heterozygosity
if the two valleys had the \emph{same} population density per unit area,
but one valley was larger than the other,
because heterozygosity reflects population size.
Or, if dispersal distances tend to be shorter
in the more densely populated valley -- 
e.g., because resources are more plentiful, 
so that individuals do not have to disperse as far to find food -- 
there may be more inbreeding over short distances,
despite higher genetic diversity within the valley as a whole.
And, of course, modern population densities may not be well-predicted by historical densities.

Migration into and out of a region presents another complication.
Roughly speaking,
geographic patterns of relatedness
are determined by a tension between population size and migration.
As discussed above,
genealogical relationships form on a time scale determined by population density,
because in a geographic region containing $N$ individuals,
the chance that two individuals are siblings is of order $1/N$.
Shared ancestry -- and thus, genetic variation -- is spatially autocorrelated
to the extent that this process of coalescence
brings lineages together more quickly than migration moves them apart,
as one traces individuals' ancestry back through the pedigree.
If the migration rate into a geographic region is much greater than $1/N$ --
i.e., if more than a few individuals in each generation had parents living outside of the region --
then the pedigree in a region cannot be considered autonomously.
In other words,
regions cannot be analyzed independently
unless a region of interest sees less than one migrant per generation
(i.e., outside the parameter regime of the structured coalescent \citep{nagylaki1998}).

\subsection{How do they move?}

Dispersal is fundamental to many questions in ecology and evolution,
such as
predicting the spread of pathogens \citep{BiekReal2010},
responses to climate change \citep{parmesan2006},
or population demography and dynamics \citep{schreiber2010interactive}.
We use the term ``dispersal'' to refer to the process that governs 
the displacement between geographic locations of parent and offspring.
For some species,
it may be feasible to directly measure dispersal 
using, e.g., telemetry or banding \citep{Cayuela2018demographic},
but this is often prohibitively expensive or difficult.
Although we usually do not know the spatial locations of ancestors in the pedigree,
the spatial locations of modern individuals can give strong clues about this.

\subsubsection{Dispersal (individual, diffusive movement; $\sigma$)}

We call the absolute value of the spatial displacement between parent and offspring locations
the \emph{dispersal distance}, and refer to it as $\sigma$.
The dispersal distance describes the quantity one could calculate by,
e.g., collecting a large number of germinating seeds
and averaging the distance from those seeds to their two parents.
It determines how quickly spatial populations mix:
larger dispersal spreads relatedness, and thus genetic variation,
throughout the species' range faster.

\begin{figure}	
    \centering
    \begin{subfigure}{0.8\textwidth}
        \centering
        \includegraphics[width=\linewidth]{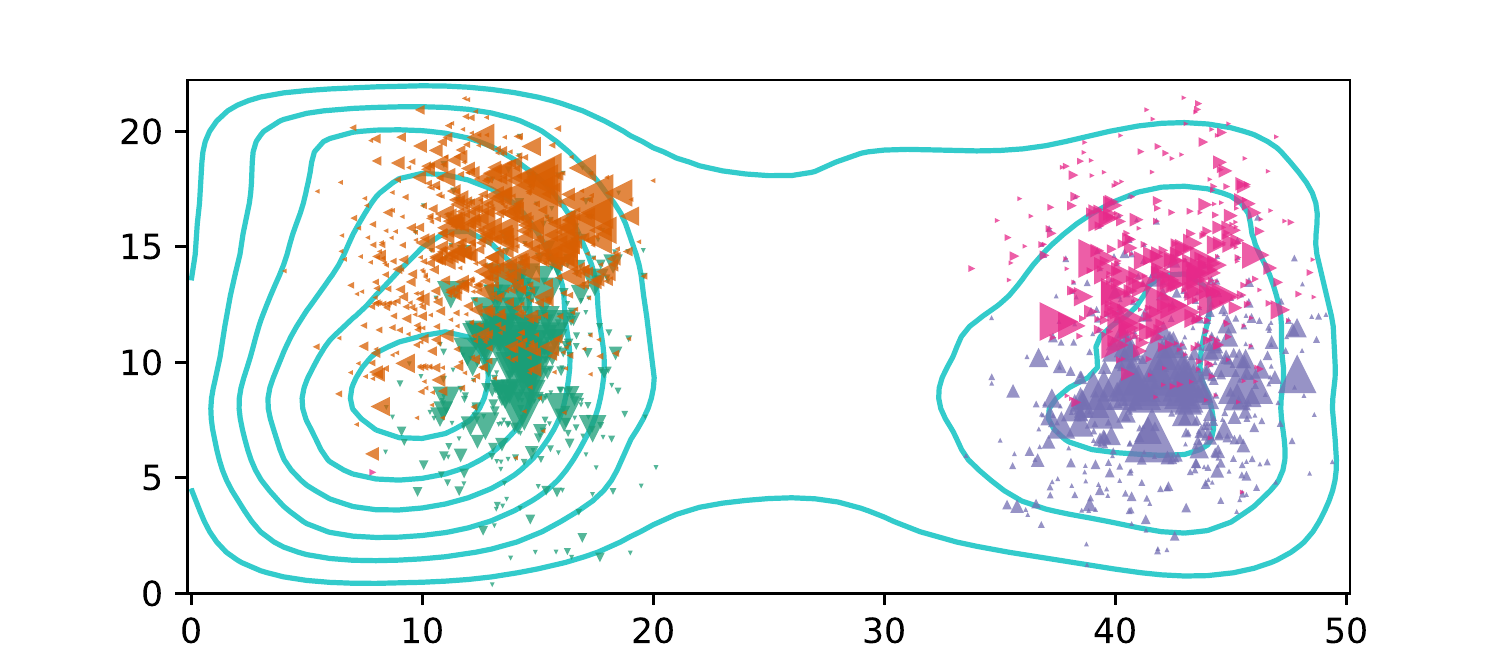}
        \caption{map of relatives}
        \label{cousin_map}
    \end{subfigure}
    \begin{subfigure}{0.7\textwidth}
        \centering
        \includegraphics[width=\linewidth]{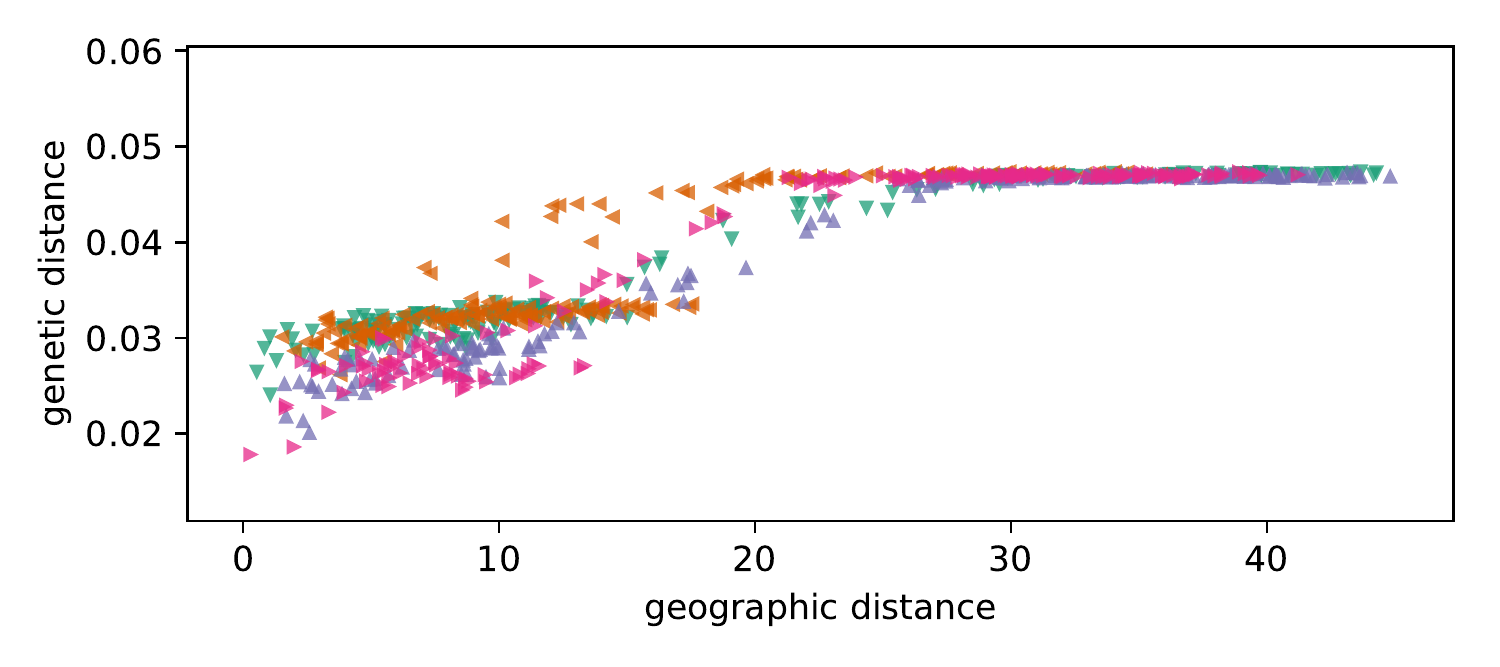}
        \caption{isolation by distance}
        \label{ibd}
    \end{subfigure}
    \begin{subfigure}{0.8\textwidth}
        \centering
        \includegraphics[width=\linewidth]{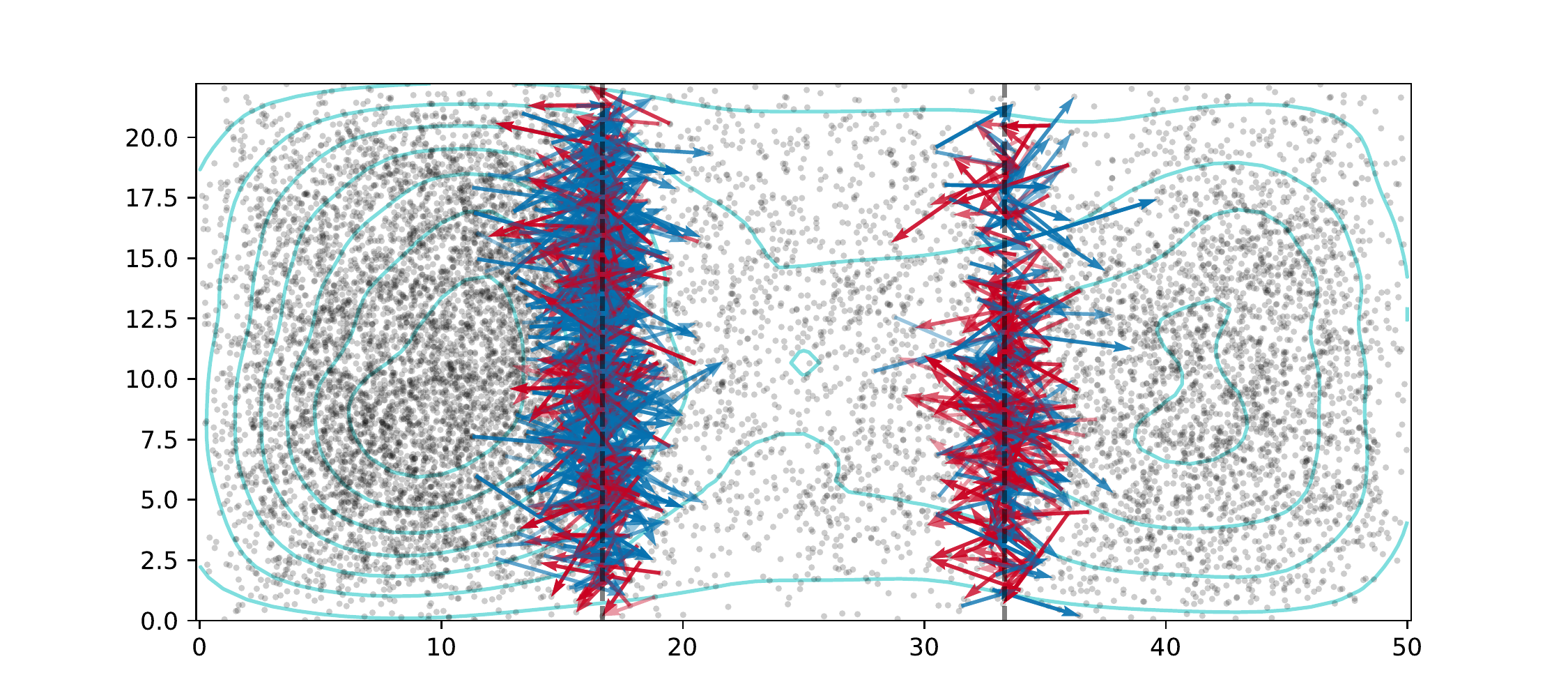}
        \caption{dispersal flux}
        \label{valleyflux}
    \end{subfigure}
        \caption{
            \textbf{(a)} Map of close genealogical relatives of four focal individuals.
            Geographic locations of 
            genealogical relatives are shown as larger triangles, with one color for each focal individual,
            with triangle size proportional to the expected proportion of genome shared
            (i.e., $2^{-k}$, where $k$ is the number of steps in the shortest path through the pedigree to the focal individual).
            \textbf{(b)}
            Genetic against geographic distance between each of four focal individuals
            and 100 other individuals, 
            from the same map, and using the same colors, as in \textbf{(a)}.
            Genetic distance (calculated as expected sequence divergence)
            increases with geographic distance within and between each valley at similar rates,
            but levels off within each valley at different values
            due to different population sizes.
            \textbf{(c)}
            Flux across two boundaries on the same continuous map as \textbf{(a)}.
            All individuals across about 8 generations are shown as small dots,
            and every parent-child relationship that crosses each of the dotted lines
            is illustrated by an arrow, colored either red (if it crosses west-to-east) or blue (east-to-west).
        }
        \label{fig:dispersal}
\end{figure}

Dispersal distance could be trivially estimated by taking
the average of these displacements across history
if we knew the spatial pedigree through time without error.
In practice, we often are only able to directly observe
the location of individuals in the present day.
However, 
spatial distances between individuals of different levels of relatedness
give us information about dispersal distance, even without parental locations.
For example,
the distance between siblings is the result of two dispersals,
so an average of this distance over many pairs of siblings 
gives an estimate of $2 \sigma$.
The mean squared displacement along
a path of total length $n$ through the pedigree is $\sigma^2 n$:
first cousins would give four times the dispersal distance, and so forth.
This spatial clustering of close relatives
is shown in Figure \ref{fig:dispersal}\subref{cousin_map}.
Unfortunately, basing an estimator of $\sigma$ on this observation
is is usually infeasible in practice,
unless a large, unbiased fraction of the population has been genotyped
\citep[e.g.,][]{Aguillon2017deconstructing}.

Empirical datasets do not usually contain many close relatives,
but using phylogenetic methods on non-recombining loci (e.g., the mitochondria),
we can obtain good estimates of times since common ancestor
between much more distantly related individuals.
This general idea was put into practice by \citet{neigel1993application}, 
who used the spatial locations of samples and their mitochondrial tree 
to infer $\sigma$ using phylogenetic comparative methods.
However, the relationship between age of common ancestor and distance apart
clearly can only apply if the geographic distance it predicts 
is much less than the width of the species range.
In many species this is not true,
because the typical time to common ancestor between two sampled individuals is long enough 
for their lineages to have crossed the species range several times \citep{barton1995genealogies}.

Instead of identifying close relatives,
we might use the decay of mean genetic relatedness with distance 
(shown in Figure \ref{fig:dispersal}\subref{ibd}) 
to learn about dispersal.
Empirical studies usually find that
the mean time since most recent common ancestor 
of two individuals at a random point on the genome
is larger for more distant individuals \citep{epperson2003geographical,charlesworth2003effects,sexton2013genetic},
and this relationship  of ``isolation by distance'' 
clearly should contain some information about $\sigma$.
An approximate formula for this relationship
in homogeneous space (Malec\'ot's formula)
can be derived in various ways 
\citep{malecot, sawyer1976branching, barton1995genealogies, rousset_1997, barton-depaulis-etheridge, robledoarnuncio2010isolation, ringbauer2017inferring, alasadi2018estimating},
but is composed of two parts: 
(a) how long has it been since the two individuals last had ancestors that lived near to each other,
and (b) how long before that did they actually share an ancestor.
The second part is not well understood,
but \citet{malecot} assumed that the second stage acts as in a randomly mating population
with some local effective size, $N_\text{loc}$.
\citet{rousset_1997} used this formula to show that
the slope of $F_{ST}$ against the logarithm of distance
is under these assumptions an estimator for Wright's neighborhood size ($4 \pi \rho_e \sigma_e^2$).
See \citet{barton1995genealogies} for more discussion of these methods.

\subsubsection{Dispersal on heterogeneous landscapes}

Of course, organisms do not live on featureless toroids, 
and individuals in one part of a species' range may disperse more than those in another 
due to heterogeneity in their nearby landscape.
In spatial population genetics, one common way 
to describe this heterogeneity is with a spatial map of ``resistance''.
The resistance value of a portion of the map is inversely proportional to the speed
that organisms tend to disperse at, or (equivalently) the distance they tend to travel,
when in that area.
These models were introduced to spatial population genetics by Brad McRae \citep{McRae2006,McRae_Beier_2007,McRae2008}, 
and represented an important step forward in modeling dispersal 
(over, e.g., least-cost path analysis).
However, resistance models do not allow 
spatial differences in dispersal variance 
or asymmetry in gene flow,
as would be seen up a fecundity gradient or following a bias in offspring dispersal downwind.
Note furthermore that we are distinguishing the ``resistance model'' of organism movement
from the (current) methods used to fit such models;
as implemented, ``resistance'' methods make use of an approximation
that can result in estimates that diverge widely from the truth,
especially if gene flow is asymmetric \citep{lundgren2018populations}.

One of the most promising current spatial methods is MAPS \citep{alasadi2018estimating}, 
which uses information about long shared haplotype tracts (recent ancestry) 
to build a map of both local population size and dispersal ability.  
MAPS builds on EEMS \citep{petkova2016visualizing}, an analogous method 
based on resistance distances.
Both methods produce spatial maps by averaging over discrete-population models
embedded in geography,
and thus share properties of both discrete and continuous population models.
In particular, the underlying dispersal model is still based on 
a migration rate (proportion of individuals replaced) between randomly mating demes.
Next we discuss a spatial analogue of this quantity.

\subsubsection{Dispersal flux}

We would often like to measure the ``amount of gene flow'' across some barrier
or between two geographic regions.
To do this,
we can ask for the number of dispersal events in a given year that cross 
a given line drawn on a map (e.g., the boundary between two adjacent regions)
-- that is, 
how many individuals are born each year whose parents
were born on the other side of the line?
This quantity, which we call the \emph{dispersal flux} across the boundary,
is useful for investigating connectivity between different areas
and the role of the landscape in shaping patterns of dispersal.
Dispersal flux is a geographical analogue of
the ``migration rate'' between discrete, isolated populations. 
Note that dispersal flux is not constrained to be equal in both directions across a boundary:
it may be easier to disperse down-current than up-current.
For instance, Figure \ref{fig:dispersal}\subref{valleyflux}
shows the number of dispersal events crossing two lines on our simulated valley landscape
over a few generations.
The line on the left goes through the denser valley, 
and so has a large flux in each direction.
The line on the right goes along the edge of the sparser valley,
and so has lower flux overall, and more east-to-west flux 
(more red than blue arrows crossing it)
due to the higher population density to the east of the line.

As before, if we had complete knowledge of the spatial pedigree,
we could calculate the flux across any line simply by counting the number of
parent-offspring pairs that span it.
However, it is rare to find any one individual
that is a direct genetic ancestor of another in a dataset
(apart from perhaps some parent-child relationships).
As a result, we almost never infer a direct chain of dispersal events,
only dispersal since coalescence events.
The likely locations of these coalescent events
depends on the maps of population density and fecundity,
thus entangling
the problems of inferring dispersal flux and population density.

For instance, across a barrier with low flux,
the average time to coalescence for pairs of individuals on opposite sides of the line
is longer than that between individuals on the same side \citep{bedassle,Duforet-Frebourg_Blum_2014,ringbauer2018estimating}.
Flux across a line is also related to mean dispersal distance, $\sigma$, discussed above.
The expected flux across a straight boundary in a homogeneous area,
per unit time and per unit of boundary length,
is equal to the birth rate per unit area multiplied by $2 \sigma / \pi$
\citep[where $\pi$ is the mathematical constant,][]{buffon1777},
so estimates of flux can be converted to estimates of dispersal distance.

\subsection{Where were their ancestors?}

Above, we have discussed how population dynamics of movement and reproduction
are reflected in the spatial pedigree.
A complementary view is to
start with individuals sampled in the present day,
then, looking backwards in time,
to ask where in space they have inherited their genomes from, 
i.e., where their ancestors lived.
Because the signal in genetic data comes from 
shared ancestry and coalescence in the pedigree,
thinking from this reverse-time perspective 
is useful in interpreting genetic data.
However, summaries of how ancestry spreads across space 
through the pedigree can be important in their own right.
For example, many humans are interested
in the locations of their own ancestors back in time.
More generally,
we are often also interested in the partitioning of genetic diversity across space,
e.g.,
quantifying the strength of inbreeding locally,
or assessing the capacity of a population to become locally adapted.
Spatial distributions of ancestry compared across sections of the genome
could also be informative about other processes, such as
the origins and mechanisms of selective sweeps.

\subsubsection{Geographic distribution of ancestry}

As one looks further back in time,
the geographic distribution of an individual's ancestors
is determined by the history of population size changes and movements.
At any point in the past,
each portion of each individual's genome can be associated 
with the ancestor from whom they inherited it.
Their genetic ancestry can therefore be apportioned across space according
to the locations of the ancestors from whom they inherited their genome.
There are a number of natural questions that arise from a description of this process.
How much of an average individual's ancestry is still within a given region, 
at some point in the past?
How long in the past was it since a typical individual in one location had an ancestor
in another geographic region (say, the next valley over)?
How far back in time do you have to go before 
the average proportion of ancestry that an individual inherits
from the local region they live in drops below (say) 90\%?

\begin{figure}	
	\begin{subfigure}{0.85\textwidth}
		\centering
	        \includegraphics[width=\linewidth]{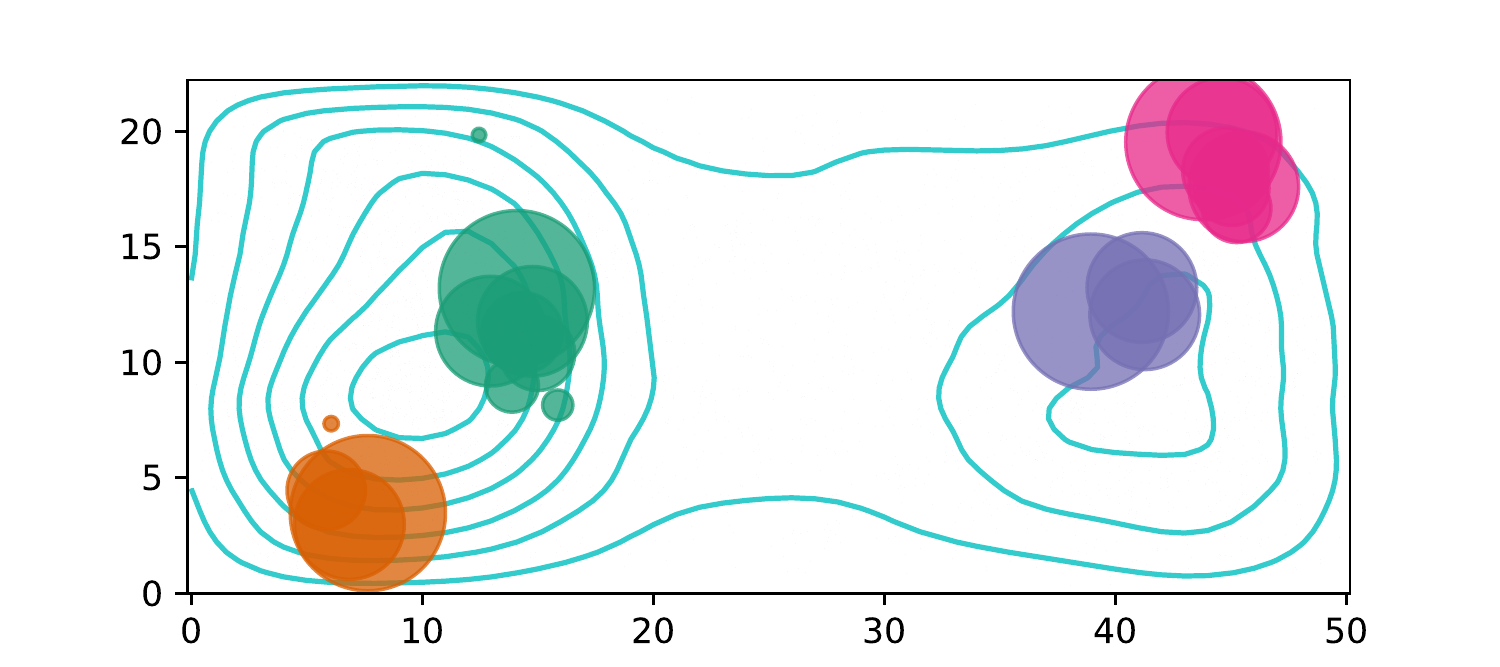}
		\caption{one generation ago}
		\label{1gen}
	\end{subfigure}\\
	\begin{subfigure}{0.85\textwidth}
		\centering
	        \includegraphics[width=\linewidth]{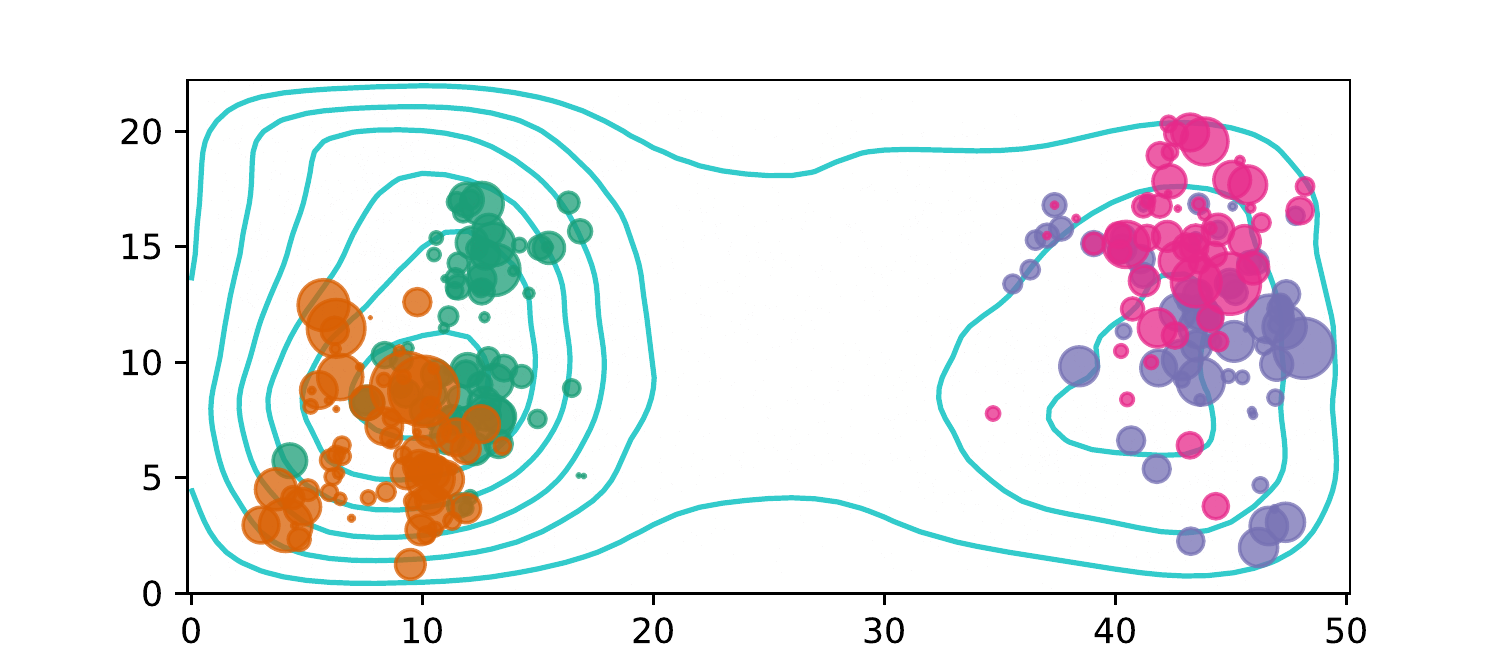}
	        \caption{sixty generations ago}
		\label{60gen}
	\end{subfigure}\\
	\begin{subfigure}{0.85\textwidth}
        		\centering
        		\includegraphics[width=\linewidth]{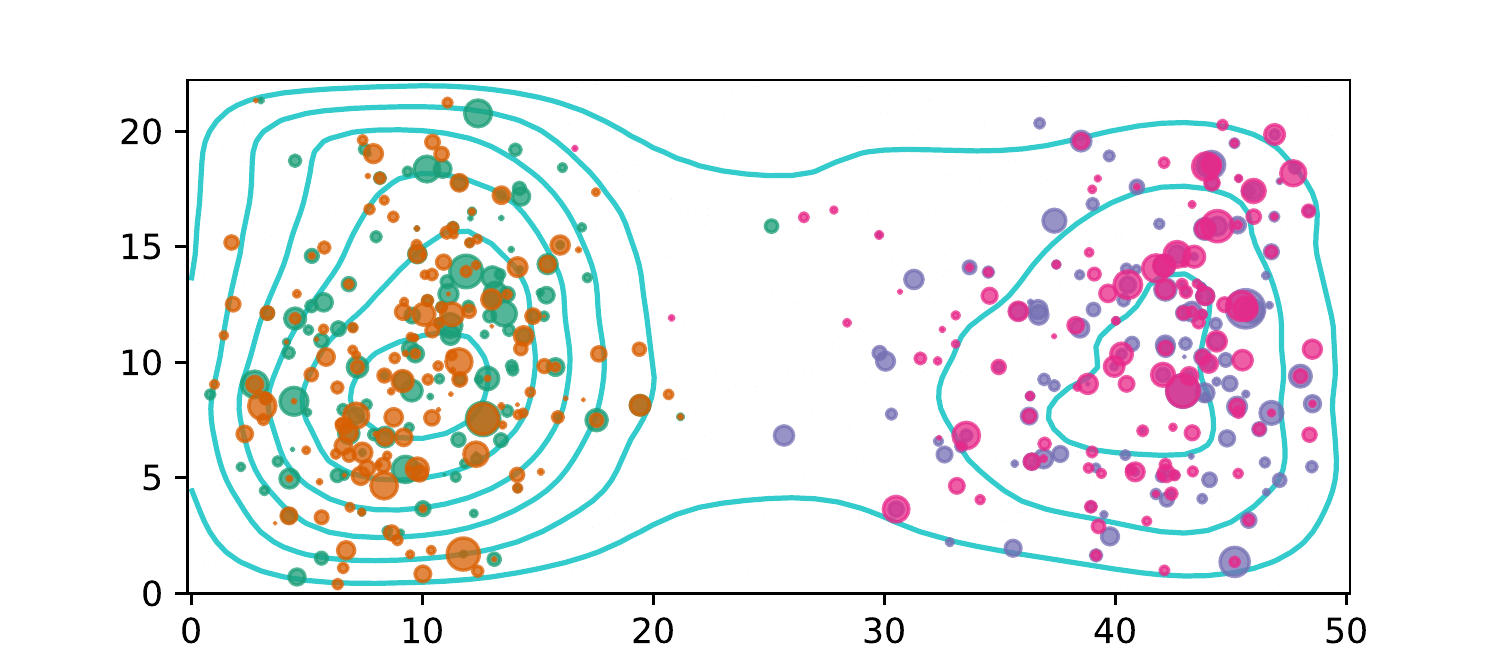}
                \caption{three hundred generations ago}
		\label{300gen}
	\end{subfigure}
        \caption{
            Spatial locations of the ancestors of four individuals:
            at 
            \textbf{(a)} one, 
            \textbf{(b)} sixty, and
            \textbf{(c)} three hundred generations in the past.
            All individuals from whom the four individuals have inherited genome are denoted by a circle,
            with one color for each individual,
            and with circle area proportional to the amount of genome inherited.
            The landscape is the same as in Figure \ref{fig:dispersal}.
            Since the simulation uses overlapping generations,
            ancestors of different degrees are present at the same time.
        }
        \label{ancestry_spread}
\end{figure}

This spatial distribution of \emph{all} genetic ancestors is depicted 
at several points back in time
for four individuals in a simulation in Figure \ref{ancestry_spread}.
As we look further back in time,
we can see the ``cloud of ancestry'' spread out across space.
The spread is diffusive,
i.e., the mean displacement of a given $n^\text{th}$-generation ancestor 
is of order $\sigma_e \sqrt{n}$,
as long as this is small relative to the width of the region.
Not shown on the map are the many genealogical ancestors
from which the focal individuals have not inherited genome --
these spread at constant speed rather than diffusively \citep{kelleher2016spread},
so that even by 15 generations in the past,
each focal individual is genealogically descended from nearly everyone alive,
even those living in the other valley \citep{chang1999}.

Although Figure \ref{ancestry_spread} does not highlight 
ancestors \emph{shared} by any pair of the focal individuals,
it is clear that these occur in the points of overlap between their two clouds of ancestry.
These shared ancestors may be informative about relative population densities in the two valleys.

\subsubsection{Relative genetic differentiation}

In the ``two valley'' simulation of Figure \ref{ancestry_spread},
ancestors are shared within valleys over the time period shown,
but not between valleys.
The difference in the geographic distribution of shared ancestors 
results in higher relatedness between the pairs of individuals 
from the same valley, 
and it underlies all measurements of relative genetic differentiation.
The most common measure of
how much more closely related neighbors are to each other relative to the population as a whole
is $F_{ST}$ \citep{Wright1951},
which can be thought of as estimating
one minus the ratio of mean coalescence time within subpopulations
to that within the entire population \citep{slatkin_1991inbreeding}.
The normalization by mean overall coalescence time 
makes this measure relative, 
and therefore comparable across populations of different sizes.

In stable populations,
relative genetic differentiation is determined by the tension between 
geographic mixing (due to dispersal flux between areas)
and local coalescence (determined by population density).
In the more remote past, each individual's ancestors are distributed across
all of geography,
so that the long-ago portion of the pedigree looks like that of a randomly mating population.
Tracing the pedigree backward in time, 
remote ancestors have ``forgotten" the geographic position of their 
modern-day descendants.
The transition between these two phases -- recent spatial autocorrelation 
(``scattering") and distant random mating (``collecting")
\citep{Wakeley1999,wilkins2004separationoftimescales}
happens on the time scale over which a lineage crosses the species range \citep{Wakeley1999}.
Since lineage motion is diffusive, in a continuous range of width $L \sigma_e$,
coalescences between lineages of distant individuals happen on a time scale of $L^2$ generations.
Relative genetic differentiation
is determined by the likelihood that coalescence between two nearby individuals
occurs during the scattering phase
-- i.e., by the length of time that two spatial clouds of ancestry of Figure \ref{ancestry_spread}
are correlated with each other,
relative to the local effective population size.
So, in a species with limited dispersal (i.e., small $\sigma_e$),
$F_{ST}$ may not be best measure of this quantity -- 
others are possible that depend less on the deep history of the population.
For instance,
one could ask for the proportion of the genomes of two neighbors 
inherited in common from ancestors in the last $T$ generations
relative to the same quantity for distant individuals,
where $T$ is chosen to reflect the time scale of interest.
The appropriate measure will depend on the application at hand.

\subsection{Clustering, grouping, and ``admixture''}

Many tools for modern analysis of genetic data
partition existing genetic variation into clusters labeled as (discrete) populations
\citep[e.g.,][]{STRUCTURE, ADMIXTURE}
The ubiquity of gene flow means that
any conceptual model of discrete populations 
must also include \emph{admixture} between them.
Models of discrete population structure and admixture 
provide visualizations that can be helpful for
describing and making sense of patterns of relatedness and genetic variation.
Coupled with a quantification of relative genetic differentiation,
they can be used in conservation to delineate discrete management units.
However, in practice, these discrete groups may in fact be determined
by clustering of sampling effort rather than any intrinsic divisions within the population.
Explicitly incorporating the geographic proximity of samples 
can help alleviate these concerns \citep{spacemix,conStruct},
but it would still be desirable to delineate groups in a way explicitly linked
to the actual history of relatedness, i.e., the spatial pedigree.

\begin{figure}	
    \centering
    	\begin{subfigure}{0.85\textwidth}
		\centering
        		\includegraphics[width=\linewidth]{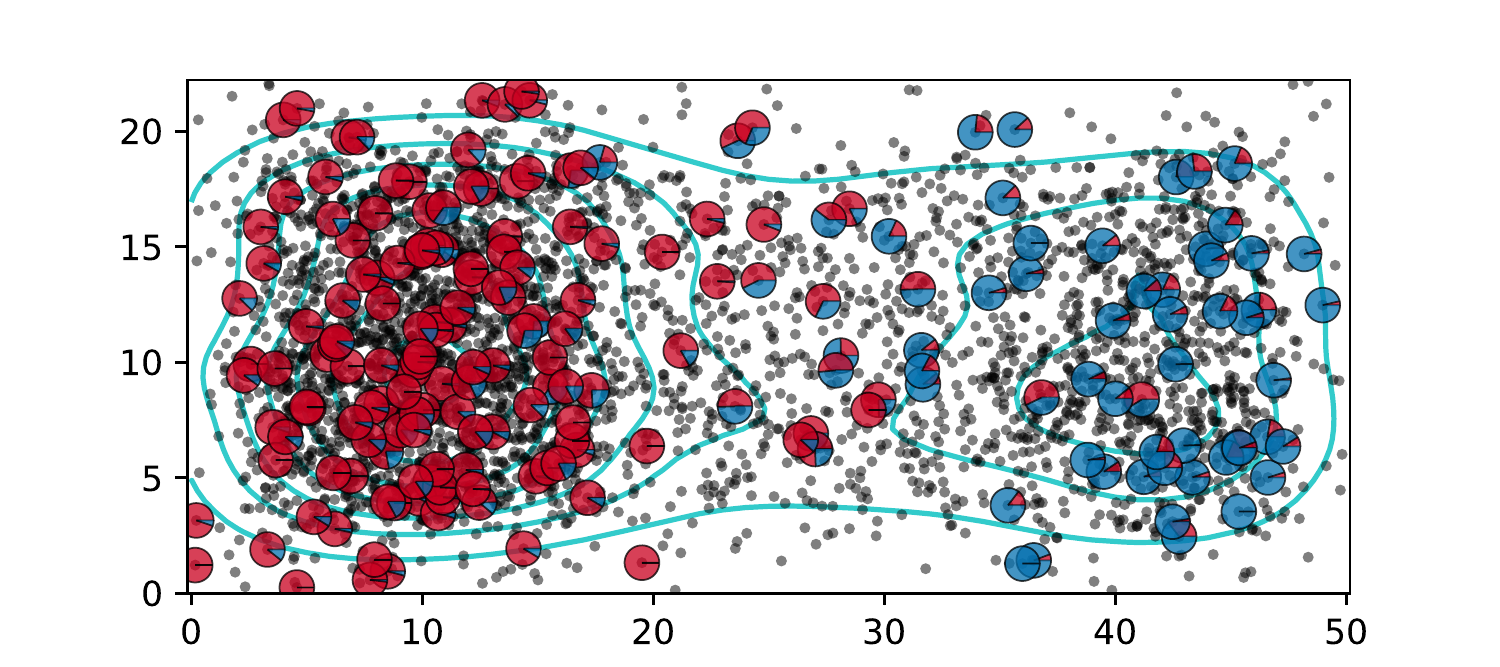}
	        \caption{eight generations post-contact}
	        \label{8gen}
    	\end{subfigure}
    	\begin{subfigure}{0.85\textwidth}
		\centering
	        \includegraphics[width=\linewidth]{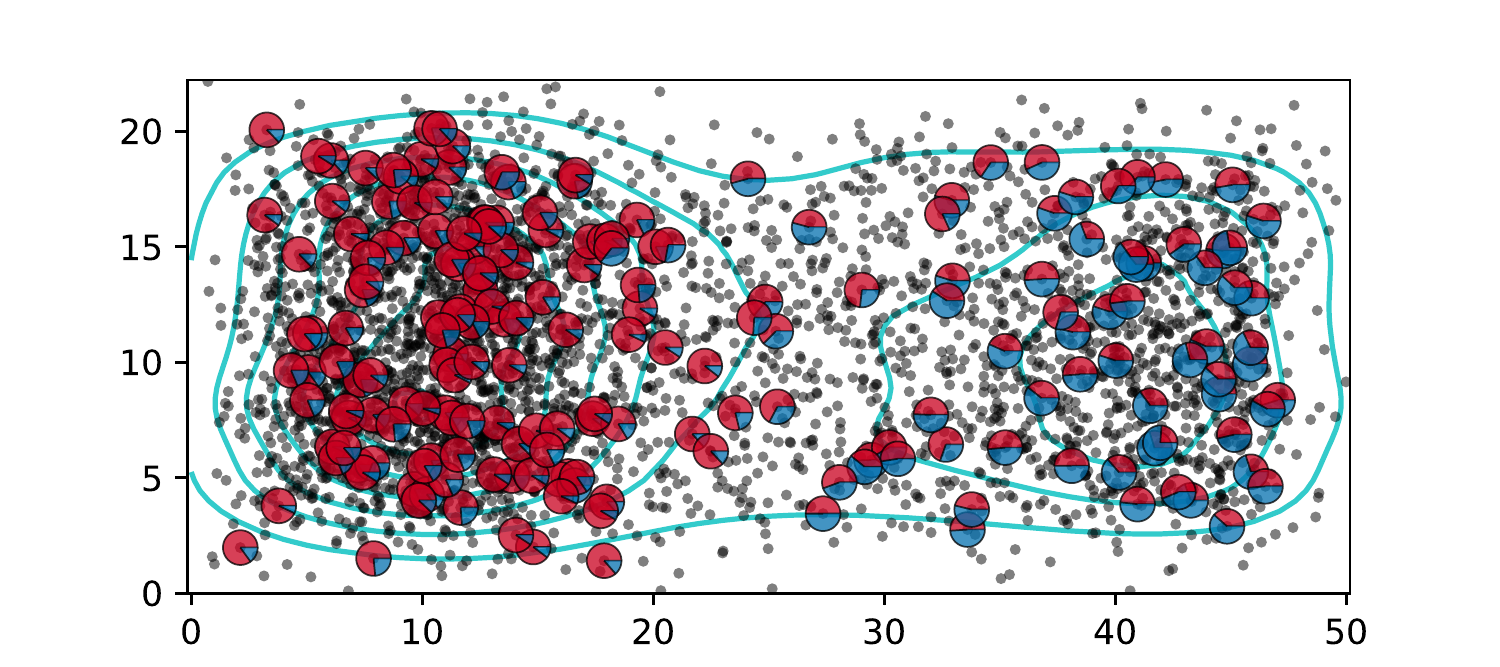}
	        \caption{seventy-five generations post-contact}
	        \label{75gen}
    	\end{subfigure}
        \caption{
            Admixture proportions after a recent secondary contact.
            Each pie shows the proportion of a diploid individual's genomes
            that inherit from the western (red) and eastern (blue) valleys, respectively.
            The plots show the geographic distribution of that ancestry 
            \textbf{(a)} Eight generations, and
            \textbf{(b)} seventy-five generations 
            after a barrier along the top of the ridge is removed.
        }
        \label{postglacial_expansion}
\end{figure}

The real-world situation in which admixture between discrete populations 
is perhaps most clearly defined is when populations that have been completely separated
for a long period of time come back into contact.
In this case, the discrete populations are defined by partitioning \emph{ancestral} individuals,
and admixture of subsequent generations is determined by inheritance from them.
For instance, suppose that 
the two valleys of our running example
were isolated during a long period of glaciation,
followed by glacial retreat and expansion into secondary contact.
The valleys clearly act as discrete populations during glaciation, 
during which there are are few if any pedigree connections between the valleys.
Pedigrees only start to overlap more recently between individuals in each valley.
To obtain an admixture proportion,
we can fix some ``reference'' time in the past,
and label the genomes of each individual in subsequent generations 
according to which of the two valleys it inherited from at the reference time.
This is shown in Figure \ref{postglacial_expansion}\subref{8gen}:
eight generations after contact,
this quantity clearly reflects the history of isolation;  
individuals in or near each valley have inherited most of their genomes 
from ancestors in the closest valley.
However, if we return 75 generations later, 
we see a more complicated story (Figure \ref{postglacial_expansion}\subref{75gen}).
The continued spatial mixing has resulted in a significant 
portion of ancestry in the west valley inherited from 
glacial-era ancestors in the east valley, 
and vice versa.
Further complicating the issue, 
the west-to-east flux across the ridge between the valleys 
is higher than that in the opposite direction 
because the western valley has a higher population density. 
As a result, a much higher proportion of the genomes
in the eastern valley are now inherited 
from individuals that, 
during glaciation, were found in the western valley.

This simulation highlights both the relevance of 
a model of discrete population structure and admixture
to spatially continuous data, 
and also its limitations.
Outside the time during and immediately following glaciation, 
describing genotyped individuals on this landscape 
using a model of discrete populations with admixture 
becomes less helpful.
Without a concrete interpretation based in the population pedigree,
partitioning genetic variation into discrete groups, 
and labeling of samples as admixed between those groups
can encourage the incorrect interpretation that discrete populations are Platonic ideals,
real and unchanging.

\subsection{How have things changed over time?}

As we have discussed above,
different methods are sensitive to signal
from different time slices of the pedigree.
Of course, the geographic distributions and dispersal patterns of most species,
not to mention the landscape itself,
have changed over the time scale spanned by the pedigree of modern individuals.
For example, 
high-latitude tree populations may only have been established
following the retreat of the glaciers
some tens of generations ago \citep{WhitlockMcCauley1999},
and invasive species may have entered their current
environment even more recently than that.
In addition to organisms moving themselves across landscapes,
landscapes are often changing out from under organisms' feet.
Anthropogenic land use
and global climate change
are radically altering both where organisms can live
and where they disperse \citep{parmesan1999}.
The result is that many empirical systems
are not in equilibrium,
so that the dynamics we estimate from
measurements of extant individuals may
not be representative of those in the past.
Exacerbating this problem,
it can take a long time to reach equilibrium
in areas of high population density
or across regions of low flux
\citep{CrowAoki1984group, whitlock1992temporal, slatkin1993isolation, WhitlockMcCauley1999}.
Theory and methods that can address change across both time and space
are still lacking,
and present an exciting set of challenges for the field.

\subsubsection{Difficult discontinuous demographies}

The specter of spatial discontinuity in the pedigree 
-- those situations in which none of the individuals living in a particular region 
at a particular point in time are descended from any of the individuals 
who had inhabited that region at some point in the past 
-- haunts the inference of historical processes from modern genetic data.
Spatial discontinuities in the pedigree occur when when regions become unoccupied
(e.g., due to ecological disturbance or agonistic interactions) 
and are recolonized by immigrants from elsewhere. 
Discontinuity is expected over long enough time scales that ranges have shifted,
and vicariance--secondary-contact events have occurred.
There is abundant evidence of large-scale population movements or changes 
in human datasets that include ancient or historical samples 
\citep{skoglund2014investigating, PickrellReich2014, lazaridis_ancient_2014, haak2015massive, joseph2018inference},
as well as in other species over recent time scales using museum specimens \citep{bi2013unlocking,bi2019chipmunks}.
Spatial discontinuity can pose a problem 
if the models we use for inference are predicated on the assumption that 
the observed geographic distribution of genetic variation 
was generated in the sampled geographic context.
In principle, we can include large-scale population movement in our inference methods,
but only if we have some clue that it has occurred.
Identifying lack of fit due to such historical events
remains an open problem.
As we develop more datasets that include genotype data from historical individuals, 
we will better be able to evaluate the prevalence of spatial discontinuity 
and its effects on our inferences across natural systems. 

\subsubsection{Genealogical strata}

The field of population genetics has many heuristics for genetic data; 
different types of genetic information are informative 
about events that occurred at different times in a population's history.
For example, patterns of sharing of rare alleles can be informative about events in the recent past,
and long shared haplotypes even more so.
In contrast, pairwise statistics like $\pi$ or $F_{ST}$ mostly inform us about events long ago,
on a scale of $N_e$ generations.
Different statistics thus provide us with windows into different strata of the pedigree.
Rates of sibling- and cousin-ship 
would shed light on the last few generations of history,
but carries the (in many cases unreasonable) requirement of a sample 
that is a sizable fraction of the entire population.
Methods for inferring the distant history of populations using a few samples
are the most well-developed \citep[e.g.,][]{dadi,Li_Durbin2011,momi},
but these can incorporate very little in terms of geographic realism,
and indeed much of the signal of geography 
may have washed out on this coalescent time scale \citep{wilkins2004separationoftimescales}.
The length spectrum of long, shared haplotypes --
also known as ``identical-by-descent" (or IBD) segments --
can be used to peer into the strata of the pedigree between these two time scales.
The lengths of segments of genome inherited in common by two individuals
from $t$ generations ago scales roughly with $1/t$, 
these can carry substantial information about ancestors from only tens of generations ago.
This fact has been used by \citet{alasadi2018estimating}
to create estimated maps of population density and migration rate
for different, recent periods of time.
However, the different strata cannot be so cleanly disentangled 
-- e.g., although 1cM--long shared haplotypes are older on average than 2cM-long ones,
the distributions of ages of these shared haplotypes overlap considerably.
Moreover, the correspondence between shared haplotype length and age of common ancestor
depends strongly on population size history \citep{ralph2013geography}.

\section{Next steps for spatial population genetics}

In this review we have focused on \emph{demographic inference} --
seeking to learn about past or present demographic processes
from population genetic data.
As demography -- birth, reproduction, death -- 
is the mechanism by which natural selection occurs,
it is fundamental not only to history and ecology but also evolutionary biology.
Describing the geographic distribution of genetic variation -- 
its patterns, 
the processes that have generated and maintained it, 
and its consequences -- 
is a fundamental goal of the field.
We have already sketched out plenty of work for the field
-- quantitative inference with continuous geography is for the most part an unsolved problem.
Nonetheless, we describe below a few more areas 
in which we hope or expect further progress in the field to be made.

\subsection{Selection}

Some of the most important aspects of geographical population genetics
we have not reviewed are related to natural selection.
A significant avenue for future work is to 
develop methods that jointly estimate selection and demography for spatially continuous populations. 
We have mostly ignored this above,
but clearly local adaptation and natural selection impact 
the structure of the pedigree, including its spatial patterns.
For example, after a hard selective sweep,
everyone in the population has inherited a single beneficial allele,
and so the lineages of all extant individuals at that locus will pass through the lucky individual
in whom that mutation occurred.
There is a great deal known about the expected geographical action of selection,
e.g., how beneficial alleles spread across space \citep{fisher1937wave,RalphCoop2010parallel,HallatschekFisher2014},
how locally adaptive alleles can be maintained in the face of gene flow \citep{slatkin1973geneflow,kruuk1999comparison,RalphCoop2015patchy},
or how hybrid zones are structured \citep{barton1985analysis,sedghifar_2015}.
In all these situations, selection can produce dramatically different patterns in the genealogies
of different regions of the genome.
Incorporating selection into spatially continuous models 
will facilitate a greater union of population genetic processes 
and ecological models of population dynamics, 
and help shed light on important evolutionary questions.

\subsection{Simulation and inference} 

Spatial population genetics suffers from a lack of concrete mathematical results
that can be used for inference.
Although we have a good understanding of the decay of expected pairwise relatedness with distance
(e.g., the Wright--Malec\'{o}t formula),
a deeper understanding of the coalescent process
is required to make quantitative predictions of more complex population genetic statistics.
How can we fill this gap?
A recent advance in the field, 
which may help improve inference models with and without selection, 
has been the advent of powerful and flexible simulation methods, 
such as SLiM \citep{haller2018forward,haller2019treesequence} and msprime \citep{kelleher2016msprime}.
These methods allow
researchers to produce genome-scale datasets from simulations of arbitrary complexity.
This capability offers an excellent pedagogical tool 
and opportunity to build intuition for patterns of genetic variation 
in complex scenarios for which theory may be lacking.
In addition, these simulation methods can be used in statistical inference.
For example, simulated datasets can be used in the training 
of machine learning methods \citep[e.g.,][]{SchriderKern2018}, 
or to compare between different generative models using 
approximate Bayesian computation \citep{MarjoramTavare2006modern}.
Simulation-based inference presents an exciting way forward 
for problems that may be analytically intractable,
although for large problems, the required computational effort is daunting.
As an intermediate solution,
a promising class of spatial coalescent models allows rapid simulation with geographically explicit models
\citep{barton2010newmodel,guindon2016demographic},
although the precise relation to forwards-time demographic models is not well understood.

\subsection{Identifiability and ill-posedness}

Another important gap in current methods for spatial population genetics inference 
is a quantitative understanding of the limits of our ability to learn about 
quantities of interest.
Demographic inference with population genetics data
is an example of an \emph{inverse problem} --
the mapping from parameters to data is relatively well-understood,
but the inverse map is not.
Many inverse problems are \emph{ill-posed},
meaning that many distinct models give rise to statistically indistinguishable datasets,
i.e., there is some degree of statistical nonidentifiability \citep{petrov2005well,stuart2010inverse}.
For example, 
as with estimating effective population size using the site frequency spectrum \citep{Myers2008},
we may never be able to infer rapid changes in effective population density
\citep[although see also][]{BhaskarSong2014descartes}.
Likewise, 
the timing of heterogeneity in flux across a particular part of the landscape 
may be difficult to disentangle from the magnitude of its change, 
as is the case with migration/isolation models 
of dynamics in discrete demes \citep{sousa2011nonidentifiability}.
More generally,
it may be difficult to learn about local biases in the direction of flux, 
or large-scale flow in the average direction of dispersal,
without genetic data from historical individuals.
Mathematical theory and simulation-based tests of inference methods
should allow us to further determine the spatial and temporal resolutions
with which we can view history.

\subsection{The past} 

The growing availability of genomic data from historical or ancient samples, 
collected from archaeological sites or museums,
is one of the most exciting developments in population genetics, 
and has the potential to revolutionize the field of spatial population genetics.
These historical data can provide a kind of ``fossil calibration,''
helping anchor spatial pedigrees in space and time.
Although it may be unlikely that any particular genotyped historical individual 
is the direct genetic ancestor of any sampled extant individual, 
the geographic position of the historical sample and its 
relatedness to modern individuals is nonetheless informative 
about the geography of any modern individual's ancestors.
These historical data can also illuminate temporal heterogeneity 
in some of the processes described in Section 3.4, 
which might otherwise be invisible in datasets comprised of only modern individuals.

\section*{Conclusion}

There is a long history of empirical research describing spatial patterns
in genetic variation \citep{dobzhansky1947,human_blood_types,landscape_genomics_review}.
Most population genetics inference methods assume a small number of randomly mating populations,
and it is still unknown what effect realistic geographies have on these inferences.
This is starting to change, however,
as larger datasets with vastly more genomic data become available,
giving us power to bring much more of a population's pedigrees into focus.
In turn, the ability to study pedigrees and their geography 
over the recent past is bringing about a union 
between the fields of population genetics and population ecology.
Measures of dispersal and density based on the pedigree 
can also inform the study of the factors governing the 
distribution and abundance of organisms 
over short timescales that might have previously been out of reach for population genetics.
As datasets, theory, and methods in spatial population genetics continue to progress, 
it will be exciting to explore further integration of the two disciplines.

\section*{Disclosure statement}
The authors are not aware of any affiliations, memberships, funding, or financial holdings 
that might be perceived as affecting the objectivity of this review.

\section*{Acknowledgments}
The authors gratefully acknowledge 
Yaniv Brandvain, Graham Coop, 
John Novembre, Doug Schemske, 
and Marjorie Weber for helpful discussions about spatial population genetics,
as well as Nick Barton, Graham Coop (again), Andy Kern, and Brad Shaffer 
for invaluable comments on the manuscript.

\bibliographystyle{ar-style1}
\bibliography{references}

\end{document}